\documentclass[12pt]{article}

\usepackage{amsmath}
\usepackage{amssymb}
\usepackage{amsfonts}
\usepackage{latexsym}

\catcode `\@=11 \@addtoreset{equation}{section}

\catcode `\@=12

\newcommand{\be}{\begin{equation}}
\newcommand{\en}{\end{equation}}
\newcommand{\bea}{\begin{eqnarray}}
\newcommand{\ena}{\end{eqnarray}}
\newcommand{\beano}{\begin{eqnarray*}}
\newcommand{\enano}{\end{eqnarray*}}
\newcommand{\bee}{\begin{enumerate}}
\newcommand{\ene}{\end{enumerate}}

\newcommand{\mc}{\mathcal}

\newcommand{\E}{{\cal E}}
\newcommand{\F}{{\cal F}}
\newcommand{\I}{{\mathbb{I}}}

\newcommand{\1}{1 \!\! 1}

\newcommand{\Hil}{\mc H}

\catcode `\@=11 \@addtoreset{equation}{section}
\catcode `\@=12

\begin{document}

\thispagestyle{empty}

\vspace*{2cm}

\begin{center}
{\Large \bf Extended SUSY quantum mechanics, intertwining operators and coherent states}   \vspace{2cm}\\

{\large F. Bagarello}\\
  Dipartimento di Metodi e Modelli Matematici,
Facolt\`a di Ingegneria,\\ Universit\`a di Palermo, I-90128  Palermo, Italy\\
e-mail: bagarell@unipa.it

\end{center}

\vspace*{2cm}

\begin{abstract}
\noindent We propose an extension of {\em supersymmetric quantum mechanics} which produces a family of isospectral hamiltonians. Our procedure slightly extends the idea of intertwining operators. Several examples of the construction are given. Further, we show how to build up vector coherent states of the Gazeau-Klauder type associated to our hamiltonians.

\end{abstract}

\vspace{2cm}


\vfill


\newpage

\section{Introduction and the method}

In some old papers the concept of {\em supersymmetric quantum mechanics} (SUSY qm) has been introduced and analyzed in many details, see \cite{suku,CKS} and references therein, and \cite{andrcan} for a more recent paper with a rather extended bibliography. The original motivation was to get a deeper insight on SUSY in the elementary particles context. In our opinion, however, the most relevant effect of this analysis was the recipe which produces a family of hamiltonians whose eigensystems can be derived by the eigensystem of a given {\em seed hamiltonian} $H_1=A^\dagger A$.
 Just to summarize those aspects of SUSY qm we are interested in, consider the self-adjoint operators $H_1=A^\dagger A$ and $H_2=AA^\dagger$, with $A$ and $A^\dagger$ not necessarily bosonic or fermionic  (in which cases the situation turns out to be not very useful!). Given now the two eigenvalue equations
$$
H_1\Psi_n^{(1)}=E_n^{(1)}\,\Psi_n^{(1)}\qquad
H_2\Psi_n^{(2)}=E_n^{(2)}\,\Psi_n^{(2)},
$$
with $E_0^{(1)}=0$ (SUSY unbroken), then the following relations can be easily proved: $E_{n+1}^{(1)}=E_n^{(2)}$, $\Psi_n^{(2)}=\frac{1}{\sqrt{E_n^{(2)}}}\,A\Psi_{n+1}^{(1)}$ and $\Psi_{n+1}^{(1)}=\frac{1}{\sqrt{E_n^{(2)}}}\,A^\dagger\Psi_{n}^{(2)}$, for all $n\geq 0$.

Now, if we put
$$
H=\left(
  \begin{array}{cc}
     H_1 & 0 \\
  0 & H_2 \\
  \end{array}
\right),\quad
Q=\left(
  \begin{array}{cc}
     0 & 0 \\
  A & 0 \\
  \end{array}
\right),\quad
Q^\dagger=\left(
  \begin{array}{cc}
     0 & A^\dagger \\
  0 & 0 \\
  \end{array}
\right),
$$
the following algebra is satisfied:
\be
[H,Q]=[H,Q^\dagger]=0,\quad Q^2=(Q^\dagger)^2=0,\quad \{Q,Q^\dagger\}=H.
\label{10}\en
This implies that if we know the eigensystem of $H_1$, $(E_n^{(1)}, \Psi_n^{(1)})$, then the eigensystem of $H_2$ can be recovered easily, and viceversa. Moreover, in a rather natural way, commutators and anti-commutators appear in the game, as in ordinary SUSY.
It should be mentioned that this procedure naturally works for $d=1$ systems, while there is not yet a generally accepted procedure for $d>1$, see for instance \cite{daspern,ioffe}.
In this paper we extend part of these results to a more general situation, where the dimensionality of the system plays no role and where it is not required to $H_1$ and $H_2$ to be factorized. The extension we propose here is rather different from that proposed in \cite{rau}, where a particular attention is given to the wave-functions rather than on the hamiltonian $H_1$, see below. As a matter of fact, our extension goes in the direction of what in literature are called {\em intertwining operators}, \cite{intop,cjt}, where two (or a family of) hamiltonians  are related to each other by some operator as in $h_1x=xh_2$, just as an example. In this case, since $x$ relates $h_1$ and $h_2$, then it is called an intertwining operator. It should be stressed that in the existing literature on this subject $h_1$, $h_2$ and $x$ are given. This, in our opinion, is a strong restriction which under suitable conditions is removed using our strategy, as it is discussed below, where we show how to construct $h_2$ once we have $h_1$ and an operator $x$ satisfying certain conditions.

We continue this section giving now the details of our method, while some examples of the construction will be discussed in  the next section, with a particular attention to an example coming from the so-called quons, see \cite{moh,fiv,gree}. In Section III we describe in some details a possible way to construct a family of vector coherent states extending the procedure originally proposed in \cite{gk}, while Section IV contains our conclusions. It may be worth stressing that, in order to make the paper more readable, we will skip most mathematical details, which mainly derive from the fact that some of the operators appearing in the game are unbounded and, as a consequence, they should be treated properly if we are interested in the mathematical rigor.

\vspace{2mm}

Let $h_1$ be a self-adjoint hamiltonian on the Hilbert space $\Hil$, $h_1=h_1^\dagger$, whose normalized eigenvectors, $\hat\varphi_n^{(1)}$, satisfy the following equation: $h_1\hat\varphi_n^{(1)}=\epsilon_n\hat\varphi_n^{(1)}$, $n\in\Bbb{N}_0:=\Bbb{N}\cup\{0\}$. Suppose that there exists an operator $x_1$ with the following properties:
\be
[x_1x_1^\dagger,h_1]=0,
\label{11}\en
and  $N_1:=x_1^\dagger\,x_1$ is invertible.  If such an operator exists and if we call
\be
h_2:=N_1^{-1}\left(x_1^\dagger\,h_1\,x_1\right),\qquad \varphi_n^{(2)}=x_1^\dagger\hat\varphi_n^{(1)}
\label{12}\en
the following conditions are satisfied:
\be \left\{
\begin{array}{ll}
\mbox{$[\alpha]$}\qquad h_2=h_2^\dagger  \\
\mbox{$[\beta]$}\qquad x_1^\dagger\left(x_1\,h_2-h_1\,x_1\right)=0\\
\mbox{$[\gamma]$}\qquad \mbox{if }\varphi_n^{(2)}\neq 0 \mbox{ then } h_2\varphi_n^{(2)}=\epsilon_n\varphi_n^{(2)}\\
\end{array}
\right. \label{13} \en
The proof of these claims is straightforward:

$[\alpha]$: we start noticing that, because of (\ref{11}) and of the associativity of the operators, which we assume here\footnote{We remind that associativity is not always granted when unbounded operators are involved, see \cite{bagrev} and references therein.},
$$
N_1\left(x_1^\dagger\,h_1\,x_1\right)=x_1^\dagger\left(x_1\,x_1^\dagger\,h_1\right)\,x_1=
x_1^\dagger\left(h_1\,x_1\,x_1^\dagger\right)\,x_1=\left(x_1^\dagger\,h_1\,x_1\right)\,N_1
$$
which implies in turns that $\left[N_1^{-1},x_1^\dagger\,h_1\,x_1\right]=0$. Therefore, since $N_1=N_1^\dagger$, property $[\alpha]$ follows.

$[\beta]$: this is a trivial consequence of the definition of $h_2$ in (\ref{12}). It is enough to multiply both sides of the first equation by $N_1$ from the left. Moreover, just taking the adjoint of the identity in $[\beta]$, we get $(h_2x_1^\dagger-x_1^\dagger h_1)x_1=0$.

Following the existing literature, \cite{intop,cjt}, condition $[\beta]$ suggests that $x_1$ is a {\em weak} intertwining operator.

$[\gamma]$: if $\varphi_n^{(2)}\neq 0$ we have
$$
h_2\,\varphi_n^{(2)}=N_1^{-1}\left(x_1^\dagger\,h_1\,x_1\right)x_1^\dagger \hat\varphi_n^{(1)}=
N_1^{-1}\,x_1^\dagger\,h_1\,\left(x_1\,x_1^\dagger\right) \hat\varphi_n^{(1)}=
N_1^{-1}\,x_1^\dagger\,\left(x_1\,x_1^\dagger\right)\,h_1\, \hat\varphi_n^{(1)}=
$$
$$
=N_1^{-1}\,N_1\,x_1^\dagger\,h_1\, \hat\varphi_n^{(1)}=x_1^\dagger\,\left(\epsilon_n\hat\varphi_n^{(1)}\right)=\epsilon_n\,\varphi_n^{(2)}.
$$
Notice that  even if $\hat\varphi_n^{(1)}$ is normalized, $\varphi_n^{(2)}$ needs not to have norm one. To avoid confusion, and since this aspect will be relevant in Section III, we will  put an {\em hat} over the normalized vectors.

\vspace{2mm}

As already mentioned the main conclusion of this approach is that within the hypothesis on $N_1$ and assuming (\ref{11}), we can define a new hamiltonian, $h_2$, whose spectrum and whose eigenvectors are related to those of $h_1$ as in SUSY qm and in the theory of intertwining operators. More in details: the spectra of the two operators coincide while the eigenvectors are related as in (\ref{12}). Moreover, if $\epsilon_n$ is not degenerate, we can also deduce that $\varphi_n^{(1)}=x_1\,\varphi_n^{(2)}$ (but, at most, for a normalization constant). Indeed from (\ref{12}) we have $x_1\,\varphi_n^{(2)}=x_1\,x_1^\dagger\,\hat\varphi_n^{(1)}$ and, using (\ref{11}),
$$
h_1\left(x_1\,\varphi_n^{(2)}\right)=h_1\left(x_1\,x_1^\dagger\right)\hat\varphi_n^{(1)}=\left(x_1\,x_1^\dagger\right)\,h_1\,
\hat\varphi_n^{(1)}=\epsilon_n\,x_1\,x_1^\dagger\,\hat\varphi_n^{(1)}=\epsilon_n\,\left(x_1\,\varphi_n^{(2)}\right),
$$
so that our claim follows.

\vspace{2mm}

{\bf Remark:--} the invertibility of $N_1$ is used in (\ref{12}) to define $h_2$. However all the examples considered in Section II show that  $h_2$ does not depend explicitly on $N_1^{-1}$, so that one may wonder whether this assumption could be avoided. However, a preliminary analysis suggests that if $N_1^{-1}$ does not exist, i.e. if zero belongs to the spectrum of $N_1$, then $h_2$ cannot be derived by $h_1$ as in (\ref{12}) (or in a similar way), and we just go back to the theory of the intertwining operators. Therefore this property seems to play a crucial role.

It should also be stressed that, but for SUSY qm, see Example 1 of the next section, the commutation rules in (\ref{10}) are not recovered in general.

\vspace{2mm}

Of course many examples fit our assumptions, and will be discussed in the next section. Before doing this, however, we want to remark that also this generalized SUSY qm gives rise to a family of {\em soluble} hamiltonians, as it happens for ordinary SUSY qm, \cite{CKS}. For that it is enough to iterate our previous procedure. Suppose therefore that an operator $x_2$ exists such that $[x_2x_2^\dagger,h_2]=0$ and  $N_2=x_2^\dagger\,x_2$ is invertible. If we now introduce $h_3=N_2^{-1}(x_2^\dagger\,h_2\,x_2)$ and $\varphi_n^{(3)}=x_2^\dagger\,\varphi_n^{(2)}=x_2^\dagger\,\left(x_1^\dagger\,\varphi_n^{(2)}\right)$, then properties $[\alpha], [\beta], [\gamma]$ are replaced by properties $
[\alpha']$:  $h_3=h_3^\dagger$, $[\beta']$: $x_2^\dagger\left(x_2\,h_3-h_2\,x_2\right)=0$ and
$[\gamma']$: if $\varphi_n^{(3)}\neq 0$ then  $h_3\varphi_n^{(3)}=\epsilon_n\varphi_n^{(3)}$. Needles to say, this procedure can be iterated as many times as we like, at least if we find operators $x_j$ having the correct properties. We will show in the next section an easy way in which these operators can be easily constructed starting from the seed hamiltonian $h_1$.

Before concluding this section, we want to point out once again the main differences between the approaches discussed here and in \cite{intop}: first of all our approach is constructive, meaning with this that once $h_1$ and $x_1$ are given, then $h_2$ can be explicitly constructed as in (\ref{12}). Secondly, all the examples considered in the existing literature, at least in our knowledge, are based on differential operators while, as it will be clear in the next section, our idea works also in different contexts, where a major relevance is given to the commutation rules. Also, not many results do exist in literature relating intertwining operators and SUSY qm to coherent states, and no result exists at all, in our knowledge, concerning coherent states of the Gazeau-Klauder type.

\section{Examples}

We will now consider some examples of our construction. As discussed above, this mainly consists in finding an hamiltonian $h$ and an operator $x$ such that $[xx^\dagger,h]=0$ and $x^\dagger\,x$ is invertible. In the first part of this section we will give some example of $h_2$ constructed from $h_1$. In the second part we will get chains of hamiltonians. It may be worth mentioning that, because many examples involving differential operators have already been  considered in the literature on intertwining operators, \cite{intop}, we focus here our interest to a different class of examples, where a key role is played by the commutation rules rather than on the (differential) expression of the operators involved.

\vspace{2mm}

{\bf Example 1.}

If $h_1=a^\dagger\,a$ and $x_1=a^\dagger$ our assumptions above are surely satisfied (at least if $[a,a^\dagger]=\1$) and we recover standard SUSY qm. Indeed, in this case we have $h_2=N_1^{-1}(a\,a^\dagger\,a\,a^\dagger)=a\,a^\dagger$, since $N_1=a\,a^\dagger\geq\1$, which is invertible even if we use the differential representation for $a=\frac{1}{\sqrt{2}}\left(x+ip\right)$.

\vspace{2mm}

{\bf Example 2.}

We consider again $h_1=a^\dagger\,a$ but we take now $x_1=(a^\dagger)^2$. Also, we assume that $[a,a^\dagger]=\1$. This assumption will be removed in the next subsection. Therefore we find $x_1\,x_1^\dagger=(a^\dagger)^2\,a^2=-N+N^2$, where $N=a^\dagger\,a=h_1$ is the standard number operator. As a consequence, $[x_1\,x_1^\dagger,h_1]=0$. Furthermore we have $N_1=x_1^\dagger\,x_1=N^2+3N+2\1\geq 2\1$, so that $N_1^{-1}$ exists independently of the representation we adopt for $a$ and $a^\dagger$. Using the commutation rule above we find that $h_2=N+2\1$, which is simply a shifted version of $N$. It is clear that $h_2$ is self-adjoint. We also find that $[\beta]$ holds in a strong form: $x_1\,h_2-h_1\,x_1=(a^\dagger)^2(N+2\1)-N(a^\dagger)^2=0$ because of the commutation rule. Let  us finally define $\hat\varphi_n^{(1)}=\frac{\left(a_1^\dagger\right)^n}{\sqrt{n!}}\,\hat\varphi_0^{(1)}$, with $a_1\hat\varphi_0^{(1)}=0$. Now we put $\varphi_n^{(2)}=x_1^\dagger\,\hat\varphi_n^{(1)}$, which is zero if $n=0,1$. If $n\geq2$ we find $\varphi_n^{(2)}=\mu_n\hat\varphi_{n-2}^{(1)}$, where $\mu_n$ is a normalization constant which could be easily computed. Therefore we get, if $n\geq2$, $h_2\varphi_n^{(2)}=(N+2\1)\varphi_n^{(2)}=(N+2\1)\mu_n\hat\varphi_{n-2}^{(1)}=\mu_n(n-2+2)
\hat\varphi_{n-2}^{(1)}=n\mu_n\hat\varphi_{n-2}^{(1)}=n\varphi_{n}^{(2)}$, as expected.

This example can be extended easily. For instance, if again $h_1=a^\dagger\,a$ but $x_1=(a^\dagger)^3$, we find $h_2=N+3\1$, and so on.

\vspace{2mm}

{\bf Example 3.}

Let $\{\epsilon_n\}$ be the eigenvalues of the hamiltonian $h_1$ and $\hat\varphi_n^{(1)}$ the related eigenvectors, $n=0,1,2,\ldots$, which span all of $\Hil$. If we call $P_n^{(1)}$ the projector operator defined as $P_n^{(1)}f=<\hat\varphi_n^{(1)},f>\,\hat\varphi_n^{(1)}$, with $n=0,1,2,\ldots$ and $f\in \Hil$ we can write $h_1=\sum_{n=0}^\infty \epsilon_n\,P_n^{(1)}$. Of course $\sum_{n=0}^\infty \,P_n^{(1)}=\1$. We introduce the operator $P_{i,j}^{(1)}$ defined on the orthonormal basis $\{\hat\varphi_n^{(1)}\}$ of  $\Hil$ as
$$P_{i,j}^{(1)}\,\hat\varphi_n^{(1)}=\left\{
\begin{array}{ll}
0 \hspace{11mm}\mbox{ if } j\neq n\\
\hat\varphi_i^{(1)}\hspace{7mm}\mbox{ if } j= n\\
\end{array}
\right.
$$
We define $x_1=\sum_{l=0}^\infty\,P_{l+1,l}^{(1)}$, which is well defined on the set of all finite linear combinations of $\{\hat\varphi_n^{(1)}\}$, which is dense in $\Hil$. Its adjoint is $x_1^\dagger=\sum_{l=0}^\infty\,P_{l,l+1}^{(1)}$.
Therefore $N_1=x_1^\dagger\,x_1=\sum_{n=0}^\infty \,P_n^{(1)}=\1$, which is clearly invertible, while $x_1\,x_1^\dagger=\1-P_0^{(1)}$ commutes trivially with $h_1$. After few computations we get $h_2=\sum_{n=0}^\infty \epsilon_{n+1}\,P_n^{(1)}$, while
$$\varphi_n^{(2)}=x_1^\dagger\,\hat\varphi_n^{(1)}=\left\{
\begin{array}{ll}
0 \hspace{13mm}\mbox{ if } n=0\\
\hat\varphi_{n-1}^{(1)}\hspace{7mm}\mbox{ if } n\geq 1\\
\end{array}
\right.
$$
and, for $n\geq 1$, $h_2\,\varphi_n^{(2)}=\epsilon_n\,\varphi_n^{(2)}$.

Once again, this example can be easily extended, taking for instance  $x_1=\sum_{l=0}^\infty\,P_{l+2,l}^{(1)}$. It is clear, however, that we get no particular insight about the eigenstates of $h_2$, since these are essentially those of $h_1$, but for a finite number.

\vspace{2mm}

We now give few examples in finite dimensional Hilbert spaces. Of course in this case using SUSY  to find the eigenstates of $h_2$ is not  necessary, since they can be computed with other methods, surely more natural. However, we sketch these examples here since they show explicitly how our procedure works and produce interesting results.

\vspace{2mm}

{\bf Example 4.}

We first consider a two-dimensional situation. The most general hamiltonian, in this case, is $h_1=\left( \begin{array}{cc}  a & c \\
 \overline{c} & b \\       \end{array}   \right)$
with $a, b\in\Bbb{R}$. We also take $x_1=\left(\begin{array}{cc}0 & \alpha \\ \beta & 0 \\ \end{array} \right)$, with $\alpha, \beta\in\Bbb{C}$. The operator $N_1=x_1^\dagger x_1=\left(\begin{array}{cc}|\beta|^2 & 0 \\ 0 & |\alpha|^2 \\ \end{array} \right)$ is invertible if and only if $\alpha$ and $\beta$ are different from zero. Moreover $[x_1\,x_1^\dagger,h_2]=0$ if $c=0$ or if $|\alpha|=|\beta|$.

If $c=0$ in $h_1$  we find $h_2=\left( \begin{array}{cc}  b & 0 \\
 0 & a \\       \end{array}   \right)$, so that the two eigenvectors $\varphi_\pm^{(2)}$ are simply proportional to  $\hat\varphi_\mp^{(1)}$.

 If $c\neq0$ and $|\alpha|=|\beta|$, we have $\alpha=|\alpha|\,e^{i\varphi_\alpha}$ and $\beta=|\alpha|\,e^{i\varphi_\beta}$. Hence $$h_2=\left( \begin{array}{cc}  b & \overline{c}\,e^{i(\varphi_\alpha-\varphi_\beta)} \\
 c\,e^{-i(\varphi_\alpha-\varphi_\beta)} & a \\       \end{array}   \right)$$ and the relation between the eigenstates of $h_1$ and $h_2$ is the one given in (\ref{12}), as the reader can easily check.

\vspace{2mm}

{\bf Example 5.}

We consider now an example arising from the theory of the angular momentum. Let $h_1=\frac{\hbar}{\sqrt{2}}\left(
                                                                                       \begin{array}{ccc}
                                                                                         0 & 1 & 0 \\
                                                                                         1 & 0 & 1 \\
                                                                                         0 & 1 & 0 \\
                                                                                       \end{array}
                                                                                     \right)
$, which is the $z$-component of $\vec J$ in a three-dimensional representation space, and let $x_1=\alpha\,\left(\begin{array}{ccc}    0 & i & 0 \\ -i & 0 & 0 \\  0 & 0 & 1 \\  \end{array}\right)=x_1^\dagger$, if $\alpha$ is real. We have $N_1=x_1^\dagger\,x_1=\alpha^2\,\1=x_1\,x_1^\dagger$. If we take $\alpha=\sqrt{2}\,\hbar$, just to continue our analogy with the angular momentum, we find that $h_2=\frac{\hbar}{\sqrt{2}}\,\left(\begin{array}{ccc}    0 & -1 & i \\ -1 & 0 & 0 \\  -i & 0 & 0 \\  \end{array}\right)$ and again its eigenstates are related to those of $h_1$  as in $\varphi_n^{(2)}=x_1^\dagger\,\hat\varphi_n^{(1)}$, $n=0,1,2$.

\vspace{3mm}

We now briefly discuss some examples giving rise to a family of hamiltonians. Let $h_1=a_1^\dagger a_1$ and $x_1=a_1^\dagger\,e^{iB_1}$, $B_1=B_1^\dagger$. Of course $x_1\,x_1^\dagger=h_1$, so that $[x_1\,x_1^\dagger,h_1]=0$. Moreover $N_1=x_1^\dagger\,x_1=a_2^\dagger\,a_2$, where $a_2=e^{-iB_1}\,a_1^\dagger\,e^{iB_1}$. Then $h_2=N_1^{-1}\,N_1^2=N_1=a_2^\dagger\,a_2$, and the following holds: $h_2=h_2^\dagger$, $x_1h_2-h_1x_1=0$, and if $\varphi_n^{(2)}=x_1^\dagger\,\hat\varphi_n^{(1)}\neq 0$ and $h_1\hat\varphi_n^{(1)}=\epsilon_n\,\hat\varphi_n^{(1)}$, then $h_2\varphi_n^{(2)}=\epsilon_n\,\varphi_n^{(2)}$. Incidentally, we see that it seems not so crucial, a posteriori, require that $N_1^{-1}$ does exist. However, we could look for $B_1$ such that this happens. We can now iterate the procedure: let $B_2=B_2^\dagger$ and $x_2=a_2^\dagger\,e^{iB_2}$. Then we find, with the same steps as before, $h_3=a_3^\dagger\,a_3$, where $a_3=e^{-iB_2}\,a_2^\dagger\,e^{iB_2}$. The analogous properties stated for $h_2$ hold for $h_3$, and for all the other $h_j$'s which are constructed in the same way. Of course, the fact that this chain of hamiltonians is generated using unitary operators, makes this construction not particularly new but, in our opinion, it shows some interesting aspects.

For instance, if $h_1=a_1^\dagger\,a_1$ with $[a_1,a_1^\dagger]=\1$ and $B_1=(a_1+a_1^\dagger)^2$, the above procedure produces $h_2=a_1\,a_1^\dagger+4(a_1+a_1^\dagger)^2+2i({a_1^\dagger}^2-a_1^2)$. If we further take $B_2=(a_2+a_2^\dagger)^2$, we get $h_3=a_1^\dagger\,a_1+16(a_1+a_1^\dagger)^2+4i({a_1^\dagger}^2-a_1^2)$, and so on. Of course, even if we have chosen here $B_1$ and $B_2$ having the same functional expression, this is just one among all the possible choices we could have done. Incidentally, we observe that in this example $[a_1,a_1^\dagger]=[a_3,a_3^\dagger]=\cdots=\1$, while $[a_2,a_2^\dagger]=[a_4,a_4^\dagger]=\cdots=-\1$.

\subsection{A chain from quons}

This subsection is entirely devoted to the construction of a chain of hamiltonians generated from the so-called quons, \cite{moh,fiv,gree}.

They are
defined essentially by their q-mutator relation \be a a^\dagger -q a^\dagger a
=\1, \qquad q\in [-1,1], \label{21} \en
between the creation and the annihilation operators $a^\dagger$
and $a$, which reduces to the canonical commutation relation for $q=1$ and to the canonical anti-commutation relation for $q=-1$. For  $q$ in
the interval $]-1,1[$, equation (\ref{21}) describes particles which are neither
bosons nor fermions. Other possible q-mutator relations have also been proposed along the years, but they will not be considered here.

In \cite{moh} it is proved that the eigenstates of $N_0=a^\dagger\,a$ are analogous to the bosonic ones, but for the normalization. More in details, if $\hat\varphi_0$ is the vacuum of $a$, $a\hat\varphi_0=0$, then $\hat\varphi_n=\frac{1}{\beta_1\cdots\beta_{n-1}}\,{a^\dagger}^n\,\hat\varphi_0$, and $N_0\hat\varphi_n=\alpha_n\hat\varphi_n$, with $\alpha_0=0$, $\alpha_1=1$ and $\alpha_n=\beta_{n-1}^2=1+q+\cdots+q^{n-1}$ for $n\geq 2$.

Let us now construct our chain extending Example 2 above. Our starting point is, as before, $h_1=a_1^\dagger\,a_1$ and $x_1=(a_1^\dagger)^2$. We assume that $a_1 a_1^\dagger -q a_1^\dagger a_1=\1$, $q\in ]-1,1[$. Then we find that $[x_1\,x_1^\dagger,h_1]=[(a_1^\dagger)^2\,a_1^2,a_1^\dagger\,a_1]=0$, and that $\tilde h_1:=x_1^\dagger h_1 x_1=N_1+qN_1a_1a_1^\dagger$, where $N_1=x_1^\dagger\,x_1=a_1^2\,{a_1^\dagger}^2$. Therefore $h_2=N_1^{-1}\tilde h_1=\1+q\,a_1\,a_1^\dagger$, which can also be written as $h_2=(1+q)\1+q^2a_1^\dagger a_1$. So we see from the first expression of $h_2$ that our procedure, but for an additive constant, exchange the positions of the creation and annihilation operators and multiply the result for the parameter $q$. We also see that $h_2$ does not depend on $N_1^{-1}$.

To construct the third hamiltonian of our chain, we consider two different choices of $x_2$. If we choose $x_{2,a}=a_1^2$, then we find out that $[x_{2,a}\,x_{2,a}^\dagger,h_2]=0$. Also, it is an easy consequence of the q-mutator relation that $\tilde h_{2,a}:=x_{2,a}^\dagger h_2 x_{2,a}=N_{2,a}\,a_1^\dagger\,a_1$, where $N_{2,a}=x_{2,a}^\dagger\,x_{2,a}$, so that $h_{3,a}=N_{2,a}^{-1}\,\tilde h_{2,a}=a_1^\dagger\,a_1$, which is exactly $h_1$. In other words, with this choice, our SUSY procedure turns out to be cyclic.

This is not the end of the story: it is possible to avoid this cyclicity  taking $x_{2,b}=(a_1^\dagger)^2$. In this case we find again that $[x_{2,b}\,x_{2,b}^\dagger,h_2]=0$ but, after few computations, we conclude that $h_{3,b}=N_{2,b}^{-1}\,\left(x_{2,b}^\dagger\, h_2\, x_{2,b}\right)=(1+q+q^2)\1+q^3\,a_1\,a_1^\dagger$ which is clearly different from $h_1$. This result can be further generalized since a comparison of $h_{3,b}$ with $h_2=(1+q)\1+q^2a_1^\dagger a_1$ shows that the next hamiltonian of this chain will be $h_4=(1+q+q^2+q^3)\1+q^4\,a_1^\dagger\,a_1$, and so on. Incidentally, we remark that none of these hamiltonians explicitly depend on $N_{2,a}^{-1}$, $N_{2,b}^{-1}$, and so on, as already stated before.

As for the eigenstates of, e.g., $h_2$, let $\hat\varphi_0^{(1)}$ be such that $a_1\hat\varphi_0^{(1)}=0$, and $\hat\varphi_n^{(1)}=\frac{1}{\beta_1\cdots\beta_{n-1}}\,{a_1^\dagger}^n\,\hat\varphi_0^{(1)}$. Then, as we have seen before, $h_1\hat\varphi_n^{(1)}=\alpha_n\hat\varphi_n^{(1)}$, with $\alpha_0=0$, $\alpha_1=1$ and $\alpha_n=\beta_{n-1}^2=1+q+\cdots+q^{n-1}$ for $n\geq 2$. For these values of $n$ we define the non-zero vectors $\varphi_n^{(2)}=x_1^\dagger\,\hat\varphi_n^{(1)}=a_1^2\,\hat\varphi_n^{(1)}=\mu_n\,\hat\varphi_{n-2}^{(1)}$, where $\mu_n$ is an un-relevant normalization constant. Then we find, for $n\geq 2$,
$$
h_2\,\varphi_n^{(2)}=\left((1+q)\1+q^2a_1^\dagger a_1\right)\left(\mu_n\,\hat\varphi_{n-2}^{(1)}\right)=\mu_n\,\left((1+q)+q^2\alpha_{n-2}\right)\,\hat\varphi_{n-2}^{(1)}=
\alpha_n\,\varphi_n^{(2)}
$$
since $(1+q)+q^2\alpha_{n-2}=1+q+q^2(1+q+\cdots+q^{n-3})=1+q+\cdots+q^{n-1}=\alpha_n$. We leave to the reader  the extension to  other hamiltonians of the chain.

\section{Gazeau-Klauder like coherent states}

In this section we will show how the framework discussed so far can be used to construct a certain type of coherent states (CS). This is a problem which has been addressed in the literature by several authors and in several different ways, see \cite{cjt,gk,alibag,fhro} and references therein. These differences arise mainly because of the non-uniqueness of the definition of what a CS should be. To be more explicit, while some author defines them as eigenvectors of some sort of annihilation operators, \cite{cjt}, someone else appears more interested in getting a resolution of the identity, \cite{gk}. In a recent paper, \cite{alibag}, the authors have constructed the so-called vector CS associated to a general SUSY hamiltonians pair, which is still another kind of CS.

Here we will consider a {\em mixed} point of view, showing how to associate to our extended SUSY system a family of vector CS which  generalizes the Gazeau-Klauder scheme, \cite{gk}, and still share with these states most of their features. Let us first recall how these states are defined and which are their main properties. These CS, labeled by two
parameters $J>0$ and $\gamma\in\Bbb{R}$, can be written in terms of the o.n.
basis of a self-adjoint operator $H=H^\dagger$, $|n>$, as \be
|J,\gamma>=N(J)^{-1}\,\sum_{n=0}^\infty\,\frac{\,J^{n/2}\,e^{-i\epsilon_n\,\gamma}}{\sqrt{\rho_n}}\,|n>,\label{41}\en
where $0=\epsilon_0<\epsilon_1<\epsilon_2<\ldots$, $\rho_n=\epsilon_n!:=\epsilon_1\cdots\epsilon_n$, $\epsilon_0!=1$,
$H|n>=\omega\,\epsilon_n\,|n>$ and
$N(J)^2=\sum_{n=0}^\infty\,\frac{\,J^{n}\,}{\rho_n}$, which converges for $0\leq J<R$, $R=\lim_n \epsilon_n$ (which could be infinite). It may be worth
noticing that the normalization of these GK-states is $N(J)^{-1}$
and formally differs from the one used in many papers on the subject.
We adopt for the moment the same notation as in the original papers. These
states satisfy the following properties:
\begin{enumerate}
\item if there exists a non negative function, $\rho(u)$, such
that $\int_0^R\,\rho(u)\,u^n\,du=\rho_n$ for all $n\geq 0$ then, introducing a
measure $d\nu(J,\gamma)=N(J)^2\,\rho(J)\,dJ\,d\nu(\gamma)$, with
$\int_{\mathbb{R}}\ldots
\,d\nu(\gamma)=\lim_{\Gamma\rightarrow\infty}\,\frac{1}{2\Gamma}\,
\int_{-\Gamma}^\Gamma\ldots\,d\gamma$, the following resolution of
the identity is satisfied: \be
\int_{C_R}\,d\nu(J,\gamma)\,|J,\gamma><J,\gamma|=\int_0^R\,N(J)^2\,\rho(J)\,dJ\,
\int_{\mathbb{R}}\,d\nu(\gamma)\,|J,\gamma><J,\gamma|=\I;
\label{42}\en \item the states $|J,\gamma>$ are {\em temporarily
stable}: \be e^{-iHt}\,|J,\gamma>=|J,\gamma+\omega t>,\quad \forall
t\in\mathbb{R};\label{43}\en \item  they
satisfy the {\em action identity}: \be
<J,\gamma|H|J,\gamma>=J\,\omega;\label{44}\en
\item they are continuous: if $(J,\gamma)\rightarrow(J_0,\gamma_0)$
then $\||J,\gamma>-|J_0,\gamma_0>\|\rightarrow 0$.
\end{enumerate}
It is interesting to observe that the states $|J,\gamma>$ are
eigenstates of the following $\gamma-$ depending annihilation-like
operator $a_\gamma$ defined on $|n>$ as follows: \be
a_\gamma\,|n>=\left\{
    \begin{array}{ll}
        0,\hspace{4.6cm}\mbox{ if } n=0,  \\
        \sqrt{\epsilon_n}\,e^{i(\epsilon_n-\epsilon_{n-1})\,\gamma}|n-1>, \hspace{0.6cm} \mbox{ if } n>0,\\
       \end{array}
        \right.
\label{45}\en whose adjoint acts as
$a_\gamma^\dagger\,|n>=\sqrt{\epsilon_{n+1}}\,e^{-i(\epsilon_{n+1}-\epsilon_{n})\,\gamma}|n+1>$.
 With standard computations
we can also check that \be a_\gamma |J,\gamma>=\sqrt{J}\,
|J,\gamma>.\label{46}\en However, it should be stressed that
$|J,\gamma>$ is not an eigenstate of $a_{\gamma'}$ if
$\gamma\neq\gamma'$.

\vspace{2mm}
The situation in the present case is a bit different. Following \cite{alibag} and using the framework introduced in Section I, we define \be
  H = \left(
        \begin{array}{cc}
          h_1 & 0 \\
          0 & h_2 \\
        \end{array}
      \right), \qquad \hat\Phi_n^{(b)}=\left(
                                     \begin{array}{c}
                                       \hat\varphi_n^{(1)} \\
                                       0 \\
                                     \end{array}
                                   \right), \quad \hat\Phi_n^{(f)}=\left(
                                     \begin{array}{c}
                                       0 \\
                                       \hat\varphi_n^{(2)} \\
                                     \end{array}
                                   \right),
                                   \label{47}\en
  where we use, following the same notation as in \cite{alibag} and with a small abuse of language, "b" for bosons and "f" for fermions.
  The set $\F=\{\hat\Phi_n^{(f)},\, \hat\Phi_n^{(b)}, \,n\geq 0\}$ forms an orthonormal basis for
the Hilbert space $\Hil_{susy}:=\Bbb{C}^2\otimes\Hil$, whose scalar product is defined as follows: given $\Gamma=\left(
                                     \begin{array}{c}
                                       \gamma^{(b)} \\
                                       \gamma^{(f)} \\
                                     \end{array}
                                   \right)$ and $\tilde\Gamma=\left(
                                     \begin{array}{c}
                                       \tilde\gamma^{(b)} \\
                                       \tilde\gamma^{(f)} \\
                                     \end{array}
                                   \right)$ then we put $<\Gamma,\tilde\Gamma>_{susy}=<\gamma^{(b)},\tilde\gamma^{(b)}>+<\gamma^{(f)},
                                   \tilde\gamma^{(f)}>$, where $<,>$ is the scalar product in $\Hil$.
                                   We also define the vectors \be
                                   \hat\Psi_n=\frac{1}{\sqrt{2}}  \left(
                                     \begin{array}{c}
                                       \hat\varphi_n^{(1)} \\
                                       \hat\varphi_n^{(2)} \\
                                     \end{array}
                                   \right)          \label{48}\en

These vectors, as well as $\hat\Phi_n^{(f)}$ and $\hat\Phi_n^{(b)}$, are normalized in $\Hil_{susy}$ and are eigenvectors of $H$ with the same eigenvalue, $\epsilon_n$. Let now $J_1$ and $J_2$ be two positive quantities and let $\gamma$ be a real variable. Let further $\delta$ be a strictly positive parameter. We now define the matrices
\be J^{1/2} = \left(
        \begin{array}{cc}
          \sqrt{J_1} & 0 \\
          0 & \sqrt{J_2} \\
        \end{array}
      \right), \qquad
E_n(\delta) = \left(
        \begin{array}{cc}
          e^{-i(\epsilon_n+\delta)\gamma} & 0 \\
          0 & e^{i(\epsilon_n+\delta)\gamma} \\
        \end{array}
      \right)
\label{49}\en
It is worth mentioning that, while $\underline J=(J_1,J_2)$ and $\gamma$ generalize $(J,\gamma)$ in (\ref{41}), $\delta$ is an extra parameter which will be useful to get a resolution of the identity for $\Hil_{susy}$. We define the following vectors:
\be
\Psi_\delta(\underline J,\gamma):=N(\underline J)^{-1/2}\sum_{n=0}^\infty\frac{J^{n/2}E_n(\delta)}{\sqrt{\epsilon_n!}}\,\hat\Psi_n,
\label{410}\en
where, as in \cite{gk}, $\epsilon_0=0$, $\epsilon_k!=\epsilon_1\cdots\epsilon_k$ and $\epsilon_0!=1$. Notice that we are adopting here the standard normalization rather than that in (\ref{41}). A different, and sometime more convenient, way to write $\Psi_\delta(\underline J,\gamma)$ is the following, as it can be easily checked:
\be
\Psi_\delta(\underline J,\gamma):=\frac{1}{\sqrt{2N(\underline J)}}\sum_{n=0}^\infty\frac{1}{\sqrt{\epsilon_n!}}\,\left(J_1^{n/2}\,e^{-i(\epsilon_n+\delta)\gamma}\,\hat\Phi_n^{(b)}+J_2^{n/2}\,e^{i(\epsilon_n+\delta)\gamma}\,
\hat\Phi_n^{(f)}\right).
\label{411}\en
The normalization $N(\underline J)$ turns out to be
\be
N(\underline J)=\frac{1}{2}(M(J_1)+M(J_2)),\label{412}\en
where $M(J):=\sum_{k=o}^\infty\,\frac{J^k}{\epsilon_k!}$, which converges for $0\leq J<R$, $R=\lim_n \epsilon_n$, which is assumed to exist (but it could be infinite).

The states $\Psi_\delta(\underline J,\gamma)$ satisfy the action identity, which now looks like
\be
<\Psi_\delta(\underline J,\gamma),H\Psi_\delta(\underline J,\gamma)>_{susy}=\frac{J_1M(J_1)+J_2M(J_2)}{M(J_1)+M(J_2)}.
\label{413}\en
Notice that this returns the standard result if we simply fix $J_1=J_2=J$. As for the temporal stability, let us define the matrix
$$
V_\delta(t) = \left(
        \begin{array}{cc}
          e^{-i(h_1+\delta)t} & 0 \\
          0 & e^{i(h_2+\delta)t} \\
        \end{array}
      \right).
$$
This matrix plays the role, in our Hilbert space $\Hil_{susy}$, of the operator $e^{-iHt}$. One can check that
\be
V_\delta(t)\Psi_\delta(\underline J,\gamma)=\Psi_\delta(\underline J,\gamma+t),
\label{414}\en
for each fixed $\delta$. This means that, independently of $\delta$, $V_\delta(t)$ leaves invariant the set of the vectors in (\ref{411}).

\vspace{2mm}

{\bf Remark:--} it should be mentioned that we are using here a little abuse of language, calling for instance {\em temporal stability} property (\ref{414}) just because it is formally identical to the property in (\ref{43}), even if it is clear that $V_\delta(t)\neq e^{\pm i(H+\delta)t}$.

\vspace{2mm}

As for the resolution of the identity, let us define a measure $d\nu(\underline J,\gamma)$ as follows: $d\nu(\underline J,\gamma)=2N(\underline J)\rho(J_1)dJ_1\,\rho(J_2)dJ_2\,d\nu(\gamma)$, where $\rho(J)$ is a function satisfying the equality $\int_0^R\,\rho(J)\,J^k\,dJ=\epsilon_k!$, \cite{gk}, $\forall k\geq 0$. The measure $d\nu(\gamma)$ is defined as above, see \cite{gk}. With these definitions it is possible to deduce that, for all fixed $\delta>0$,
\be
\int_\E \,d\nu(\underline J,\gamma)\,|\Psi_\delta(\underline J,\gamma)><\Psi_\delta(\underline J,\gamma)|=\1,
\label{415}\en
where $\E=\{(\underline J,\gamma): 0\leq J_1<R,\,0\leq J_2<R,\,\gamma\in\Bbb{R}\}$.

\vspace{2mm}

{\bf Remark:--} It is worth noticing that if we take $\delta=0$ in this computation we would conclude that $\int_\E \,d\nu(\underline J,\gamma)\,|\Psi_\delta(\underline J,\gamma)><\Psi_\delta(\underline J,\gamma)|=\1+|\hat\Phi_0^{(b)}><\hat\Phi_0^{(f)}|+|\hat\Phi_0^{(f)}><\hat\Phi_0^{(b)}|$, which is not what we want. This is why $\delta$ was originally introduced into the game. The computations above are straightforward consequences of the expression for $d\nu(\gamma)$ and simply mean  that the integral $\int_\E \,d\nu(\underline J,\gamma)\,|\Psi_\delta(\underline J,\gamma)><\Psi_\delta(\underline J,\gamma)|$ is not uniformly continuous in $\delta$. We also want to mention that, instead of introducing the extra parameter $\delta$, we could as well replace $\epsilon_n$ with $\epsilon_n-\epsilon_0$ in the definition of $\epsilon_n!$ above. We prefer our approach since it adds an extra freedom to the CS we are constructing.

\vspace{2mm}

The analogy between the $\Psi_\delta(\underline J,\gamma)$'s and the Gazeau-Klauder CS can be pushed further. Indeed, let us define on the orthonormal basis $\F$ of $\Hil_{susy}$ the following $\gamma-$depending operator:
\be
A_\gamma\hat\Phi_n^{(b)}=\left\{
\begin{array}{ll}
0 \hspace{40mm}\mbox{ if }  n=0\\
\sqrt{\epsilon_n}\,e^{i(\epsilon_n-\epsilon_{n-1})\gamma}\hat\Phi_{n-1}^{(b)}\hspace{7mm}\mbox{ if } n\geq 1\\
\end{array}
\right.\label{416}\en
and
\be
A_\gamma\hat\Phi_n^{(f)}=\left\{
\begin{array}{ll}
0 \hspace{40mm}\mbox{ if }  n=0\\
\sqrt{\epsilon_n}\,e^{-i(\epsilon_n-\epsilon_{n-1})\gamma}\hat\Phi_{n-1}^{(f)}\hspace{5mm}\mbox{ if } n\geq 1\\
\end{array}
\right.\label{417}\en
Then the adjoint $A^\dagger_\gamma$ satisfies the following:
\be
\left\{
\begin{array}{ll}
A_\gamma^\dagger\hat\Phi_n^{(b)}=
\sqrt{\epsilon_{n+1}}\,e^{-i(\epsilon_{n+1}-\epsilon_{n})\gamma}\hat\Phi_{n+1}^{(b)}\\
A_\gamma^\dagger\hat\Phi_n^{(f)}=
\sqrt{\epsilon_{n+1}}\,e^{i(\epsilon_{n+1}-\epsilon_{n})\gamma}\hat\Phi_{n+1}^{(f)}\\
\end{array}
\right.\label{418}\en
Then the states $\Psi_\delta(\underline J,\gamma)$ are {\em eigenstates} of the operator $A_\gamma$ in the following sense:
\be
A_\gamma\Psi_\delta(\underline J,\gamma)=J^{1/2}\Psi_\delta(\underline J,\gamma)\label{419}\en
for all fixed $\delta$.

We conclude the section showing how to relate the operators $A_\gamma$ and $A_\gamma^\dagger$ with the  operator $x_1$ in (\ref{11}), at least in some particular situations. For that let us define
$$
L= \left(
        \begin{array}{cc}
          0 & 0 \\
          x_1^\dagger & 0 \\
        \end{array}
      \right),\qquad  L^\dagger= \left(
        \begin{array}{cc}
          0 & x_1 \\
          0 & 0 \\
        \end{array}
      \right)\qquad \mbox{and}\qquad X=L+L^\dagger=\left(
        \begin{array}{cc}
          0 & x_1 \\
          x_1^\dagger & 0 \\
        \end{array}
      \right)
$$
Then we have $L\hat\Phi_n^{(b)}=X\hat\Phi_n^{(b)}=\alpha_n^{(1)}\hat\Phi_n^{(f)}$ and $L^\dagger\hat\Phi_n^{(f)}=X\hat\Phi_n^{(f)}=\alpha_n^{(2)}\hat\Phi_n^{(b)}$, where $\alpha_n^{(1)}=\|x_1^\dagger\hat\varphi_n^{(1)}\|$ and $\alpha_n^{(2)}=\|x_1\hat\varphi_n^{(2)}\|$. Here we are assuming that $h_1$ has a non-degenerate spectrum.

If we now compute the action of $X$ on $\Psi_\delta(\underline J,\gamma)$ we easily find
$$
X\,\Psi_\delta(\underline J,\gamma)=\frac{1}{\sqrt{2N(\underline J)}}\sum_{n=0}^\infty\frac{1}{\sqrt{\epsilon_n!}}\,\left(J_2^{n/2}\,e^{i(\epsilon_n+\delta)\gamma}\,\alpha_n^{(2)}\,
\hat\Phi_n^{(b)}+J_1^{n/2}\,e^{-i(\epsilon_n+\delta)\gamma}\,\alpha_n^{(1)}\,
\hat\Phi_n^{(f)}\right).
$$
Now, if we just have $\alpha_n^{(1)}=\alpha_n^{(2)}=:\alpha$ for all $n\geq0$, which is the case if, for instance, $x_1$ is an unitary operator, then we deduce that
\be
X\,\Psi_\delta(\underline J,\gamma)=\alpha\,\Psi_\delta(\tilde{\underline J},-\gamma),
\label{420}\en
where $\tilde{\underline J}=(J_2,J_1)$.

Maybe more interesting is the situation in which $\alpha_n^{(1)}=\alpha_n^{(2)}=\epsilon_n$. In this case, after few computation, we deduce that
\be
X\,\Psi_\delta(\underline J,\gamma)=A_{-\gamma}^\dagger\,\tilde J^{1/2}\,\Psi_\delta(\tilde{\underline J},-\gamma),
\label{421}\en
where $\tilde J^{1/2}$ coincides with the matrix $J^{1/2}$, but with $J_1$ and $J_2$ exchanged. Notice that equation (\ref{421}) relates the intertwining operators $x_1$ and $x_1^\dagger$ with the CS-operator $A_{-\gamma}^\dagger$, which is exactly what we wanted to do.

\section{Conclusions}

In this paper we have considered an extended view of SUSY qm which is indeed very close to the idea of intertwining operators. We have discussed several examples of our framework, constructing pairs of isospectral hamiltonians starting from few ingredients and adopting a purely operatorial point of view. In the second part of the paper we have shown how this extended SUSY can be naturally related to vector CS of the Gazeau-Klauder type satisfying all the {\em classical} requirements of CS.

\end{document}